\documentclass{elsart}
\usepackage{graphicx}
\usepackage{color}
\usepackage{epsfig}

\begin{document}
\begin{frontmatter}
\title{Relaxation dynamics of the Kuramoto model with uniformly
distributed natural frequencies}
\author{Anandamohan Ghosh$^1$} and
\author{Shamik Gupta$^2$} 

\address{1. Indian Institute of Science Education and Research-Kolkata, Mohanpur 741252, India}
\address{2. Laboratoire de Physique Th\'{e}orique et Mod\`{e}les
Statistiques, UMR 8626, Universit\'{e} Paris-Sud 11 and CNRS,
B\^{a}timent 100, Orsay F-91405, France}

\date{\today}

\thanks[ananda]{E-mail: anandamohan@iiserkol.ac.in}
\thanks[shamik]{E-mail: shamikg1@gmail.com}
\begin{abstract}
The Kuramoto model describes a system of globally coupled
phase-only oscillators with distributed natural frequencies. The model
in the steady state exhibits a phase transition as a function of the
coupling strength, 
between a low-coupling incoherent phase in which the oscillators
oscillate independently and a high-coupling
synchronized phase.
Here, we consider a uniform distribution for the natural 
frequencies, for which the phase transition is known to be of first order. We
study how the system close to the phase transition in the supercritical
regime relaxes in time to the steady state while starting from an
initial incoherent state. In this case, numerical simulations of finite
systems have demonstrated that the relaxation occurs as a
step-like jump in the order parameter from the initial to the
final steady state value, hinting at the existence of metastable states.
We provide numerical evidence 
to suggest that the observed metastability is a finite-size effect,
becoming an increasingly rare event with increasing system size.
\end{abstract}

\end{frontmatter}
\newcommand{\gae}{\lower 2pt \hbox{$\, \buildrel {\scriptstyle >}\over {\scriptstyle
\sim}\,$}}
\newcommand{\lae}{\lower 2pt \hbox{$\, \buildrel {\scriptstyle <}\over {\scriptstyle
\sim}\,$}}

\def\be{\begin{equation}}
\def\ee{\end{equation}}
\def\bea{\begin{eqnarray}}
\def\eea{\end{eqnarray}}
\def\bef{\begin{figure}}
\def\eef{\end{figure}}
\def\bml{\begin{mathletters}}
\def\eml{\end{mathletters}}
\def\l{\label}
\def\om{\omega}
\def\O{\Omega}
\def\b{\bullet}
\def\no{\nonumber}
\def\fr{\frac}
\def\th{\theta}
Coupled oscillators that have their natural
frequencies distributed according to a given distribution, for
example, a Gaussian, a Lorentzian, or a uniform distribution, often
exhibit collective synchronization in which a finite fraction of
the oscillators oscillates
with a common frequency. Examples include groups
of fireflies flashing in unison \cite{Buck:1976,Buck:1988}, networks of pacemaker cells in
the heart \cite{Peskin:1975,Michaels:1987},
superconducting Josephson junctions
\cite{Wiesenfeld:1996,Wiesenfeld:1998}, and many others. Understanding
the nature and emergence of synchronization from the
underlying dynamics of such systems is an
issue of great interest. A paradigmatic
model in this area is the so-called Kuramoto model involving
globally-coupled oscillators \cite{Kuramoto:1975}. Although studied
extensively in the past, the model continues to raise new questions,
and has been a subject of active research; for reviews, see
\cite{Strogatz:2000,Acebron:2005}.

One issue that has been explored in recent years, and is also the focus
of this paper, concerns the Kuramoto model with uniformly distributed natural frequencies.
In this case, it is known that in the limit of infinite system size,
where size refers to the number of oscillators, the
system in the steady state undergoes a first-order phase transition across a critical
coupling threshold $K_c$, from a low-coupling incoherent
phase to a high-coupling synchronized phase. For values of the coupling
constant slightly higher than $K_c$, non-trivial relaxation dynamics
has been reported, based on numerical simulations of
large systems \cite{Pluchino:2006,Miritello:2009}. Namely, it has been shown that  
initial incoherent states while evolving in time get stuck in metastable
states before attaining synchronized steady states. This phenomenon
has been demonstrated by the
temporal behavior of the order parameter characterizing the phase
transition, which shows a relaxation from the initial
value of the order parameter to its final steady state value in step-like jumps. 
An aspect of the Kuramoto model which is of interest and
has been explored in some detail concerns finite-size effects
\cite{Daido90,Hildebrand07}, which may have important consequences, for
example, for $K < K_c$, in stabilizing the incoherent state which in the
limit of infinite size is known to be linearly neutrally stable \cite{Buice07}.
In this context, it is important to investigate whether the metastable
states mentioned above may be attributed to finite-size effects.
In this paper, we systematically study this phenomenon of step-like
relaxation. We provide numerical evidence to suggest that
the observed metastability is indeed a finite-size effect,
becoming an increasingly rare event with increasing system size.

The Kuramoto model consists of $N$ phase-only oscillators labeled by the index
$i=1,2,\ldots,N$. Each oscillator has its own natural
frequency $\om_i$ distributed according to a given probability density
$g(\om)$, and is coupled to all the other oscillators. The phase of the oscillators evolves in time according to \cite{Kuramoto:1975} 
\be
\fr{d\th_i}{dt}=\om_i+\fr{K}{N}\sum_{j=1}^N\sin(\th_j-\th_i),
\l{timeevolution}
\ee
where $\th_i$, the phase of the $i$th oscillator, is a periodic variable
of period $2\pi$, and
$K \ge 0$ is the coupling constant. 

The Kuramoto model has been mostly studied for a unimodal
$g(\om)$, i.e., one which is symmetric about
the mean frequency $\om=\O$, and which decreases monotonically and
continuously to zero with increasing $|\om-\O|$ \cite{Strogatz:2000,Acebron:2005}. Then, it is known that in the
limit $N \to \infty$, the system of oscillators in the steady state undergoes
a continuous transition at the critical threshold $K_c=2/\pi
g(0)$. For $K<K_c$, each
oscillator tends to oscillate independently with its own natural
frequency. On the other hand, for $K > K_c$, the
coupling synchronizes the phases of the oscillators, and in the limit $K
\to \infty$, they all oscillate with the mean frequency $\O$. 
The degree of synchronization in the system at time $t$ is 
measured by the complex order parameter
\be
{\bf r}(t)=r(t)e^{i\psi(t)}=\fr{1}{N}\sum_{j=1}^N e^{i\th_j(t)},
\ee
with magnitude $r(t)$ and phase $\psi(t)$, in terms of which the time evolution (\ref{timeevolution}) may be written as
\be
\fr{d\th_i}{dt}=\omega_i+Kr(t)\sin(\psi(t)-\th_i).
\l{timeevolution1}
\ee
Here $r(t)$ with $0 \le r(t) \le 1$ measures the phase coherence of the
oscillators, while $\psi(t)$ gives the average phase.
When $K$ is smaller than $K_c$, the quantity $r(t)$ while starting from any initial
value relaxes in the long-time limit to zero, corresponding to an
incoherent phase in the steady state. For $K >
K_c$, on the other hand, $r(t)$ grows in time to asymptotically saturate
to a non-zero steady state value $r_{\rm st}=r_{\rm st}(K) \le 1$ that increases continuously
with $K$.
The relaxation of $r(t)$ to steady state is exponentially
fast for $K > K_c$. For $K < K_c$, however, the nature of relaxation
depends on $g(\om)$. When $g(\om)$ has a compact support, $r(t)$ while starting from
any initial value decays to zero more slowly than any exponential as
$t \to \infty$ \cite{Strogatz:1992}. When $g(\om)$ is supported on the
whole real line, $r(t)$ as a function of time is known only in
particular cases. For example, for a Lorentzian $g(\om)$,
and a specific
initial condition, $r(t)$ decays exponentially to zero
\cite{Strogatz:1992}. For
other choices of $g(\om)$ in this class and for other initial conditions, the dependence of 
$r(t)$ on time is not known analytically, and it has been speculated that
$r(t)$ is a sum of
decaying exponentials \cite{Strogatz:1992}. 

In the limit $N \to \infty$, the state of the oscillator system at
time $t$ is described by the probability distribution $f(\th,t,\om)$
that gives for each natural frequency $\om$ the fraction of
oscillators with phase $\th$ at time $t$.
The time evolution of
$f(\th,t,\om)$ satisfies the continuity equation for the conservation
of the number of oscillators with natural frequency $\om$, and is given by a
non-linear partial integro-differential equation \cite{Strogatz:2000}.
Recent analytical studies for a unimodal $g(\om)$ (specifically, a
Lorentzian) and for two different bimodal $g(\om)$'s (given by a
suitably defined sum and difference of two Lorentzians) demonstrated by considering a restricted class of
$f(\th,t,\om)$, and by employing an ansatz due to Ott and Antonsen that the time
evolution in terms of the integro-differential equation may be exactly
reduced to that of a small number of ordinary differential
equations (ODEs) \cite{Ott:2008,Martens:2009,Pazo:2009}. 
Interestingly, the ODEs for the reduced system
contain the whole spectrum of dynamical behavior of the full system.
The Ott-Antonsen ansatz has also been
applied to various globally and nonlinearly coupled oscillators with uniformly 
distributed frequencies \cite{Baibalatov:2010}.

A uniform $g(\om)$ with a compact support does not qualify as a unimodal distribution. In
this case, it is known that in the limit $N \to \infty$, the Kuramoto
model in the steady state exhibits a first-order phase transition 
between an incoherent and a synchronized phase at the critical coupling
$K_c =2/\pi g(0)$ \cite{Pazo:2005}. 
For large $N$, numerical studies of the relaxation of an initial state with uniformly
distributed phases have demonstrated that for values of $K$
around $K_c$ in the supercritical regime, the process occurs as a step-like jump in $r(t)$ from its initial to the steady state value. 
One may interpret this behavior as suggesting the existence of metastable states in the system
\cite{Pluchino:2006,Miritello:2009}.
Our motivation is to investigate the implication of the existence of
the step-like relaxation, and whether such relaxation can be seen only in finite-sized systems.

\begin{figure}[h!]
\begin{center}
\includegraphics[width=94mm]{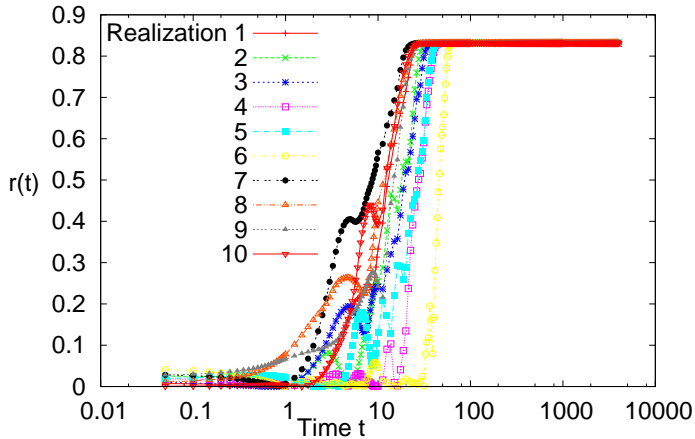}
\caption{For a given realization of the $\omega_i$'s, independently and
uniformly distributed in $[-2,2]$, the figure shows
the time evolution of the $r(t)$ for different realizations of the
$\theta_i$'s,  independently and
uniformly distributed in $[-\pi,\pi]$. Here, the system size is
$N=1000$, while the coupling constant is $K=2.62$, the critical coupling
being $K_c \approx 2.55$.}
\label{fig1}
\end{center}
\end{figure}

We performed extensive numerical simulations involving integration of Eq.
(\ref{timeevolution1}) by a $4$th-order Runge-Kutta algorithm. We
considered a system of $N=1000$ oscillators, with the $\omega_i$'s independently and uniformly distributed
in $[-2,2]$, so that $g(\omega)=1/4$ for $\omega \in [-2,2]$, and is
zero otherwise. We chose the initial state to be homogeneous in phases:
$\theta_i$'s are independently and uniformly distributed in $[-\pi,\pi]$. We took $K=2.62$, the
critical value being $K_c=2/\pi g(0)\approx 2.55$. In
simulations, we monitored the evolution of $r(t)$ in time. To discuss
our results, let us note that in the Kuramoto model, one
simulation run of the dynamics differs from another in that it corresponds to (i) a
different realization of the set of initial phases $\{\th_i\}_{1 \le i
\le N}$, and (ii) a different
realization of the set of natural frequencies
$\{\omega_i\}_{1 \le i \le N}$. In order to distinguish between the effects of the two, we
performed two sets of simulations, by fixing the initial
$\theta_i$'s and running simulations for different
realizations of the $\omega_i$'s, and vice versa.

(a) For a given realization of the $\omega_i$'s, Fig. \ref{fig1} shows $r(t)$ as a function of
time for different realizations of the initial $\theta_i$'s.
We see that in all cases, $r(t)$ shows similar relaxation behavior,
jumping in a step-like manner from the initial to the final steady state value corresponding to a synchronized phase.

\begin{figure}[h!]
\begin{center}
\includegraphics[width=94mm]{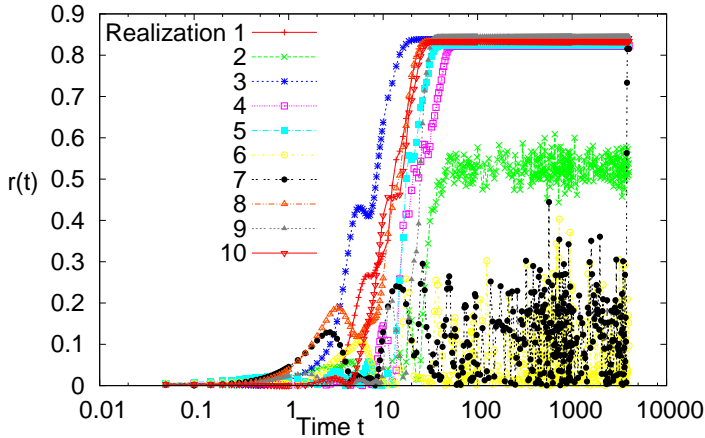}
\caption{For a given realization of the $\theta_i$'s, independently and
uniformly distributed in $[-\pi,\pi]$, the figure shows
the time evolution of the $r(t)$ for different realizations of the
$\omega_i$'s,  independently and
uniformly distributed in $[-2,2]$. Here, the system size is
$N=1000$, while the coupling constant is $K=2.62$, the critical coupling
being $K_c \approx 2.55$.}
\label{fig2}
\end{center}
\end{figure}

(b) In the second set of simulations, we focussed on different realizations
of the frequency distribution, while keeping the set of initial
$\theta_i$'s fixed.
In Fig. \ref{fig2}, we see that, similar to Fig. \ref{fig1}, the
quantity $r(t)$ jumps in a step from the initial to the final steady state value corresponding to
a synchronized phase. However, an important difference is that across realizations, there is a wide
range of values of the jump time or the relaxation time $\tau$. For
realization 2, the
system settles into a partially synchronized state. However, we
have checked that this state is not the true steady state, the latter
being a synchronized state, the relaxation to which takes place at very
long times. Thus,
the partially synchronized state may be interpreted as only a metastable
state.  
From the above numerical experiment, it is clear that the occurrence of the metastable state is dependent on the realization
of the frequency distribution only, as we had kept the initial $\theta_i$'s
fixed in the experiment. 

\begin{figure}[h!]
\begin{center}
\includegraphics[width=94mm]{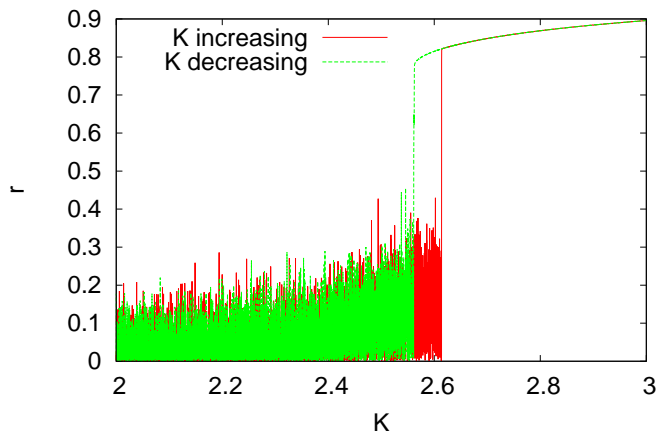}
\caption{Starting from an initial homogeneous state, the figure shows
$r$ as a function of adiabatically and cyclically tuned $K$ for the
realization $5$ in Fig. \ref{fig2}.} 
\label{fig3}
\end{center}
\end{figure}

\begin{figure}[h!]
\begin{center}
\includegraphics[width=94mm]{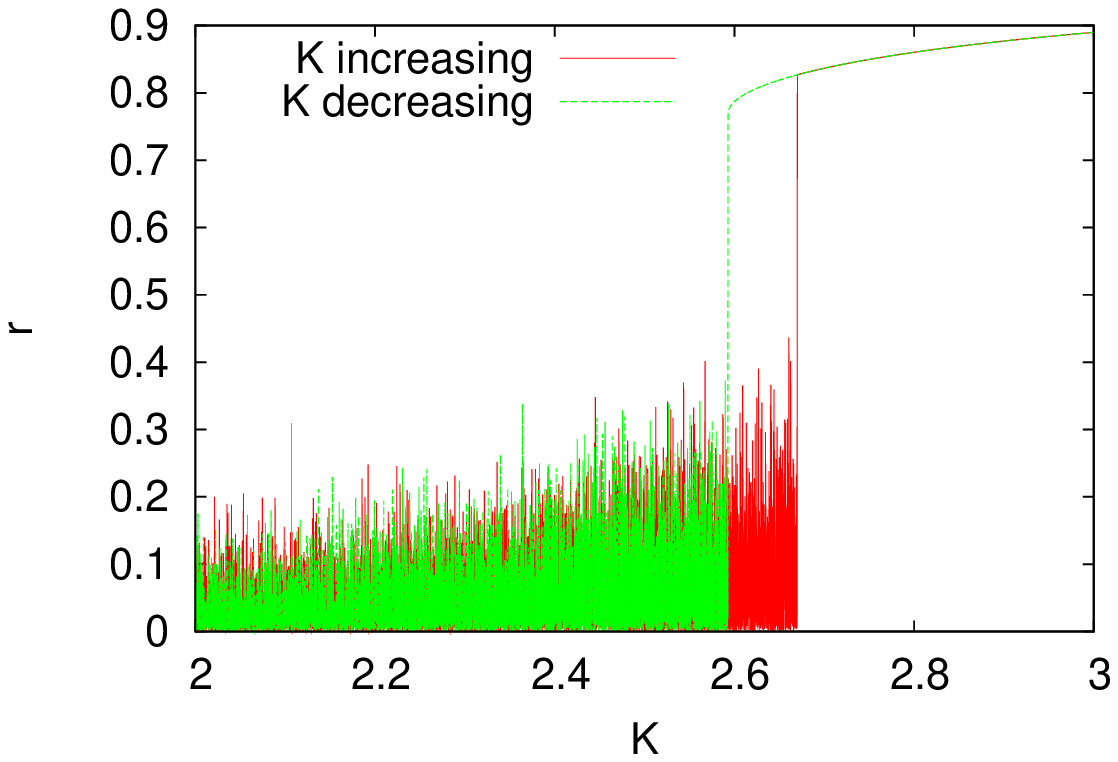}
\caption{Starting from an initial homogeneous state, the figure shows
$r$ as a function of adiabatically and cyclically tuned $K$ for the
realization $6$ in Fig. \ref{fig2}.} 
\label{fig4}
\end{center}
\end{figure}

Let us mention a different way to demonstrate the reluctance of some frequency
realizations to relax to a synchronized state, as observed in Fig.
\ref{fig2}, by performing the following numerical experiment. From the
realizations depicted in Fig.
\ref{fig2} that relax to the synchronized state within the time duration
shown, the
typical relaxation time $\tau_{\rm typical}$ may be estimated. In simulations, we prepared the system in a
homogeneous state, as in Fig. \ref{fig2}, and tuned the coupling
parameter $K$ cyclically, from low to high values and back, while
simultaneously measuring the order parameter $r$. We tuned $K$ at a rate
much smaller than the inverse of
$\tau_{\rm typical}$, thereby ensuring that the system during the course
of tuning of $K$ remains close to the instantaneous steady state at all
times. The result is the hysteresis loop depicted
in Figs. \ref{fig3} and \ref{fig4}. For one of the realizations depicted in Fig.
\ref{fig2} that is relaxing within the time duration shown, say, the 
realization $5$, Fig. \ref{fig3} shows the corresponding plot of $r$ vs.
$K$, illustrating hysteretic behavior. Such a behavior is expected of a
first-order transition, as is the case here in the model with uniformly
distributed frequencies. One may observe from the hysteresis plot that the value of $K$ used in
Fig. \ref{fig2}, namely, $K=2.62$, is well outside the hysteresis loop where the only stable
state is the synchronized state. This is consistent with Fig.
\ref{fig2}. For a realization in Fig.
\ref{fig2} that is not relaxing within the time duration shown, say, the
realization $6$, the value $K=2.62$ lies within the corresponding
hysteresis loop displayed in Fig. \ref{fig4}. The latter shows that at
this value of $K$, the stable state
of the system is
indeed not a synchronized state, but a non-synchronized state,
consistent with Fig. \ref{fig2}. However, note that for this
realization, the actual relaxation time evident from Fig. \ref{fig2} is
actually much larger than $\tau_{\rm typical}$, thus demanding that $K$ be
tuned at a slower rate than the one used to obtain Fig. \ref{fig4}.

\begin{figure}[h!]
\begin{center}
\includegraphics[width=94mm]{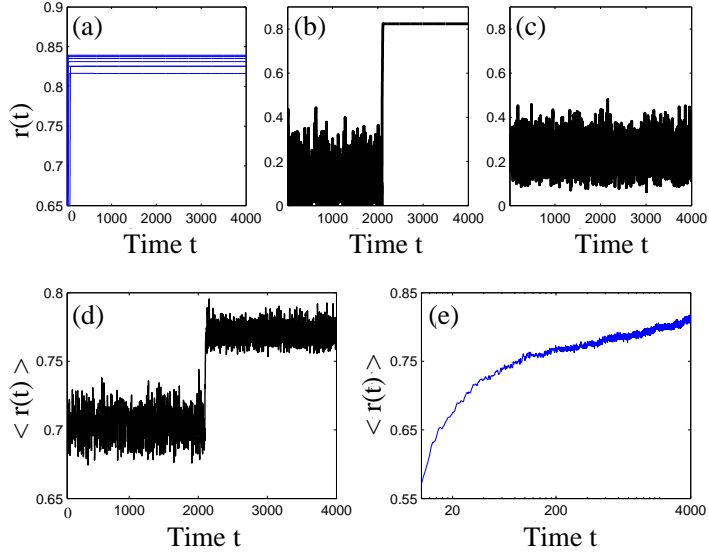}
\caption{(a) Evolution of $r(t)$ in time for $8$ different realizations
of the initial homogeneous state. Here, $N=1000, K =2.62$, while the critical coupling is $K_c \approx
2.55$. The plot shows that for these realizations, $r(t)$ jumps in a step at times $\sim 100$ from its initial
to its steady state value. (b) and (c) also show the time
evolution of $r(t)$ for two other realizations of the initial state: for the realization in (b), $r(t)$ shows a
jump to the steady state value at time $\approx 2000$, while for the one in (c), $r(t)$ remains close to its initial value and does
not relax to the steady state value in the time shown. (d) and (e) show
respectively the time evolution of $r(t)$ when averaged over the
realizations in (a)-(c), and over typical $1000$ realizations.}  
\label{fig5}
\end{center}
\end{figure}

\begin{figure}[h!]
\begin{center}
\includegraphics[width=88mm]{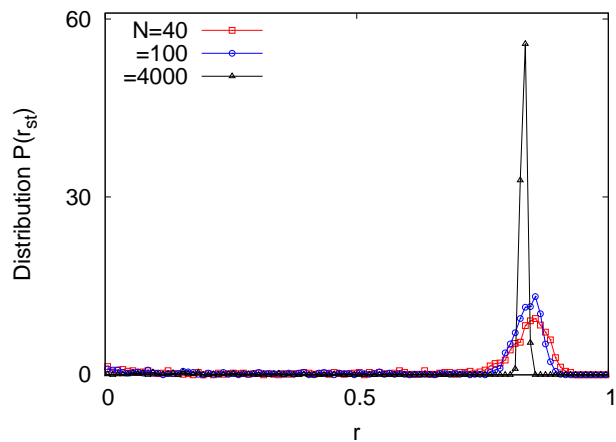}    
\caption{Starting from an initial homogeneous state, the figure shows
the distribution $P(r_{\rm st})$ of the steady state value $r_{\rm st}$ of the order
parameter for $K=2.62$.} 
\label{fig6}
\end{center}
\end{figure}

An average of $r(t)$ over
different realizations of the $\omega_i$'s shows the following
behavior. Figure \ref{fig5}(a) shows the temporal evolution of
$r(t)$ for 8 realizations with different $\omega_i$'s,
illustrating that $r(t)$ jumps in a step at times $\sim 100$ from its initial
to its steady state value of $\approx 1$ (with fluctuations that 
decrease with increasing system size $N$). For
the realization shown in Fig. \ref{fig5}(b), however, the jump to the steady
state value takes place at time $\approx 2000$, while for the realization
in Fig. \ref{fig5}(c), $r(t)$ stays close to the
initial value, and the jump does not take place in the time
duration shown. 
The average of $r(t)$ over the realizations in (a)-(c) is
shown in Fig. \ref{fig5}(d), which exhibits a step-like
relaxation, and may be interpreted as suggesting the existence of
metastable states. Similar behavior of the average $r(t)$, where the averaging
is over a {\em few} realizations, was reported in \cite{Pluchino:2006}. 
Figure \ref{fig5}(e) shows the time evolution of $r(t)$ when averaged over a
{\em large} ($\sim 1000$) number of realizations, which however does not show any
step-like relaxation. Now, as $N$ tends to $\infty$, it is reasonable to
expect that the system becomes self-averaging. This observation,
combined with our result that $r(t)$ when
averaged over a large number of realizations of $\omega_i$'s for a finite system does
not show any step-like relaxation, suggests that in the
thermodynamic limit, the relaxation of $r(t)$ occurs as a smooth process. Indeed, the probability distribution of the steady state value $r_{\rm st}$ of the order
parameter, displayed in Figure \ref{fig6}, shows that with increasing system size,
the distribution is more and more peaked at a value close to one, while the
probability to obtain any other value gets negligibly smaller. The form of
$P(r_{\rm st})$ suggests that with increasing $N$, metastable states
during the relaxation occur with increasing rarity, so that in the
thermodynamic limit, relaxation of $r(t)$ is a smooth process; had the 
occurrence of metastable states not been rarer with increasing $N$, one
would not have
had $P(r_{\rm st})$ peaked at a single value of $r_{\rm st}$. 

\begin{figure}[h!]
\begin{center}
\includegraphics[width=88mm]{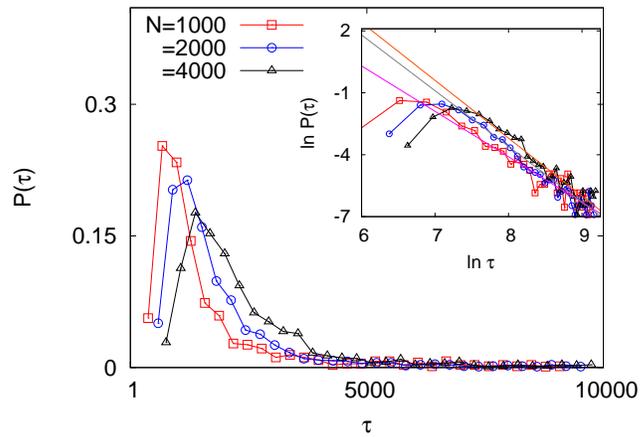}    
\caption{Distribution $P(\tau)$ of the relaxation time $\tau$. Here,
$K=2.62$. The inset shows power-law fit to the tail
of $P(\tau)$ according to Eq. (\ref{taudist}), giving values
$2.2$, $2.7$, and $2.8$ for system sizes $N=1000$, $2000$ and $4000$, respectively.} 
\label{fig7}
\end{center}
\end{figure}

\begin{figure}
\begin{center}
\includegraphics[ width=88mm]{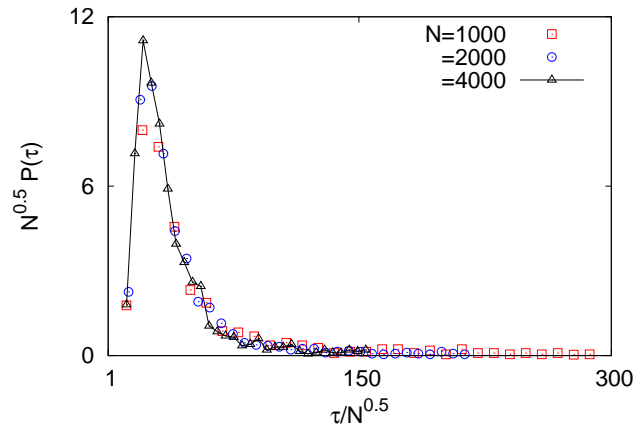}
\caption{Scaling collapse of the data for the distribution $P(\tau)$ shown in Fig.
\ref{fig7}.} 
\label{fig8}
\end{center}
\end{figure}

It is worthwhile to study the distribution $P(\tau)$ of the jump time
$\tau \ge 1$.
Our numerical data, displayed in Fig. \ref{fig7}, show that $P(\tau)$ has a power-law
tail with an exponent greater than $1$; the scaling collapse of the data, shown in Fig. \ref{fig8},
suggests the following scaling form:
\be
P(\tau) \sim \sqrt{N}f\left(\frac{\tau }{\sqrt{N}} \right),
\l{Ptau-scaling-form}
\ee
where $f(x)$ is the scaling function.

\begin{figure}
\begin{center}
\includegraphics[width=95mm]{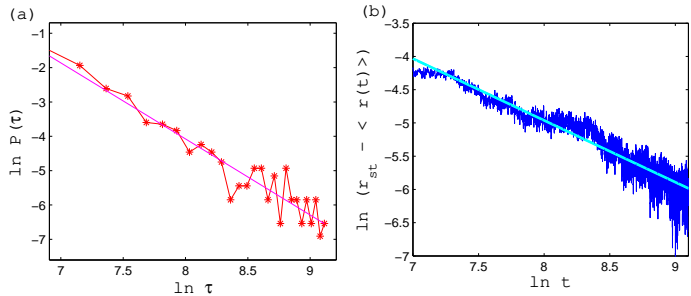}
\caption{(a) Power-law fit to the tail of $P(\tau)$, according to Eq.
(\ref{taudist}). (b) Power-law fit to the quantity
$r_{\rm st}-\langle r \rangle(t)$, according to Eq.
(\ref{order-parameter-prediction}).  Here, $N=1000, K=2.62$, while the
averaging is over $\sim 1000$ realizations. } 
\label{fig9}
\end{center}
\end{figure}

In the limit of large $N$, and in the power-law regime of $P(\tau)$, let us write
\be
P(\tau) = A\tau^{-\alpha}, 
\label{taudist}
\ee
where $\alpha > 1$ is the decay exponent, and $A$ is a constant. In the long-time regime, 
$r(t)$ will have the steady state value $r_{\rm st}
\approx 1$ (the exact value of $r_{\rm st}$ may be estimated by using results from
\cite{Pazo:2005}), provided that the jump from the initial close-to-zero value
has already taken place by this time; otherwise, $r(t) \approx 0$.
It then follows that the average order parameter, $\langle r
\rangle(t)$, for large $t$ will be
\bea
\langle r \rangle(t) &=& r_{\rm st} \int_1^t d\tau~P(\tau) \nonumber \\
&=&r_{\rm
st}\Big[\int_1^\infty d\tau~P(\tau)-\int_t^\infty d\tau~P(\tau)\Big].
\l{rteqn}
\eea
Note that in writing down Eq. (\ref{rteqn}), we made use of the fact
that the fluctuations in the steady state value $r_{\rm st}$ in the
limit of large $N$ are negligibly small.
Now, $P(\tau)$ being normalized, the first integral on the right hand
side of Eq. (\ref{rteqn}) equals one. Moreover, since $t$ is large, the second integral may be 
evaluated by using the form (\ref{taudist}) for $P(\tau)$. We get
\be
\langle r \rangle(t)=r_{\rm st}\Big[1-\fr{A}{\alpha-1}t^{1-\alpha}\Big],
\ee
where we have used the fact that $\alpha > 1$. It then follows that
\be
\ln(r_{\rm st}-\langle r \rangle(t))=\ln \Big(\fr{r_{\rm
st}A}{\alpha-1}\Big)-\beta \ln t,
\l{order-parameter-prediction}
\ee
with
\be
\beta=\alpha-1.
\l{order-parameter-prediction-1}
\ee
The prediction of Eq. (\ref{order-parameter-prediction}) is easily checked from our simulation results.
Figure \ref{fig9}(a) shows the power-law fit to the tail of $P(\tau)$ for
$N=1000$, giving $\alpha \approx 2.2$.
On the other hand, Fig. \ref{fig9}(b) shows the power-law fit to the
quantity $r_{\rm st}-\langle r \rangle
(t)$ at long times, giving $\beta \approx 0.9$ in Eq.
(\ref{order-parameter-prediction}), while the
value predicted by Eq. (\ref{order-parameter-prediction-1}) is $1.2$.
One may hope to obtain a closer agreement with better statistics and
larger $N$.

To summarize and conclude, we considered the Kuramoto model with
uniformly distributed frequencies, and studied the relaxation of a
homogeneous non-steady state in time to the steady state for values of $K$ close to the 
phase transition ($K \gae K_c$). Our numerical simulations for finite
systems showed that for fixed initial phases, but for different
realizations of the natural frequencies, the order parameter relaxes as a
step-like jump from its initial ($\approx 0$) to its steady
state value ($\approx 1$), and that there is a wide range of values of the
relaxation time $\tau$ across realizations. We demonstrated that the
distribution $P(\tau)$ has a power-law tail. In a finite system, averaging $r(t)$
over a few realizations naturally shows a 
step-like relaxation but which disappears when averaged over a
sufficiently large number of realizations. In the thermodynamic limit, when the system
becomes self-averaging, our observation that $r(t)$ when
averaged over a large number of realizations for a finite system does
not show any step-like relaxation suggests that in this limit, the relaxation of $r(t)$ occurs as a smooth process. 
The observed metastability is more a finite-size effect, its
occurrence being an increasingly rare event with increasing $N$. 
Our results on the distribution
$P(r_{\rm st})$ of the steady state order parameter clearly show that
with the increase of system size $N$, the distribution gets sharply
peaked around a single value close to $1$. 
This is in conformity with our observation that metastable states become
rarer with increasing $N$.
In order to answer quantitatively the issues raised in this work, a rigorous mathematical analysis of the relaxation
dynamics of the Kuramoto model with uniform frequency distribution
is very much desirable.

\section*{Acknowledgments}
SG acknowledges the Indo-French Centre for the Promotion of Advanced
Research under Project 4604-3 and the contract ANR-10-CEXC-010-01 for
support. SG acknowledges useful discussions with Stefano Ruffo and
Hiroki Ohta. 


\begin{thebibliography}{99}
\bibitem{Buck:1976} J. Buck and E. Buck, Sci. Am. {\bf 234}, 74 (1976).
\bibitem{Buck:1988} J. Buck, Quart. Rev. Biol. {\bf 63}, 265 (1988).
\bibitem{Peskin:1975}C. S. Peskin, Math. Aspects of Heart Physiology, Courant Institute of Mathematical Science Publications, New York, 1975, pp. 268-278.
\bibitem{Michaels:1987}D. C. Michaels, E. P. Matyas, and J. Jalife, Circulations Res. {\bf 61}, 704 (1987).
\bibitem{Wiesenfeld:1996} K. Wiesenfeld, P. Colet, and S. H. Strogatz, Phys. Rev. Lett. {\bf 76}, 404 (1996).
\bibitem{Wiesenfeld:1998} K. Wiesenfeld, P. Colet, and S. H. Strogatz, Phys. Rev. E {\bf 57}, 1563 (1998).
\bibitem{Kuramoto:1975} Y. Kuramoto, in {\em International Symposium of
Mathematical Problems in Theoretical Physics}, edited by H. Araki,
Lecture Notes in Physics Vol. 39 (Springer, Berlin, 1975).
\bibitem{Strogatz:2000}S. H. Strogatz, Physica D {\bf 143}, 1 (2000).
\bibitem{Acebron:2005}J. A. Acebr\'{o}n, L. L. Bonilla, C. J. P\'{e}rez Vicente,
F. Ritort, and R. Spigler, Rev. Mod. Phys. {\bf 77}, 137 (2005).
\bibitem{Pluchino:2006}A. Pluchino and A. Rapisarda, Physica A {\bf 365}
184 (2006).
\bibitem{Miritello:2009}G. Miritello, A. Pluchino, and A. Rapisarda,
Europhys. Lett. {\bf 85}, 10007 (2009).
\bibitem{Daido90}H. Daido, J. Stat. Phys. {\bf 60}, 753 (1990).
\bibitem{Hildebrand07}E. Hildebrand, M. A. Buice, and C. C. Chow,  
Phys. Rev. Lett. {\bf 98}, 054101 (2007).
\bibitem{Buice07} M. A. Buice and C. C. Chow, Phys. Rev. E {\bf 76},
031118 (2007).
\bibitem{Strogatz:1992}S. H. Strogatz, R. E. Mirollo, and P. C. Matthews,
Phys. Rev. Lett. {\bf 68}, 2730 (1992).
\bibitem{Ott:2008}E. Ott and T. A. Antonsen, Chaos {\bf 18}, 037113
(2008).
\bibitem{Martens:2009}E. A. Martens, E. Barreto, S. H. Strogatz, E. Ott,
P. So, and T. M. Antonsen, Phys. Rev. E {\bf 79}, 026204 (2009).
\bibitem{Pazo:2009}D. Paz\'{o} and E. Montbrio, Phys. Rev. E {\bf 80},
046215 (2009). 
\bibitem{Baibalatov:2010} Y. Baibalatov, M. Rosenblum, Z. Zh. Zhanabaev and A. Pikovsky,
Phys. Rev. E {\bf 82}, 016212 (2010).
\bibitem{Pazo:2005}D. Paz\'{o}, Phys. Rev. E {\bf 72}, 046211 (2005); L.
Basnarkov and V. Urumov, Phys. Rev. E {\bf 76}, 057201 (2007); Phys.
Rev. E {\bf 78}, 011113 (2008).
\end{thebibliography}
\end{document}